\documentclass[a4paper]{jpconf}

\newcommand{\mb}[1]{\mathbf{#1}}
\newcommand{\bds}[1]{\boldsymbol{#1}}

\newcommand{\lra}[1]{\langle #1 \rangle }

\def\be{\begin{equation}}
\def\ee{\end{equation}}

\usepackage{amsmath,amssymb}
\usepackage{graphicx}
\usepackage{natbib}
\begin{document}
\title{The FDF or LES/PDF method for turbulent two-phase flows}

\author{S.~Chibbaro$^1$, Jean-Pierre Minier$^2$ }

\address{$^1$ Institut Jean Le Rond D'Alembert 
University Pierre et Marie Curie et CNRS UMR7190, 
4, place Jussieu 75252 Paris Cedex 05}
\address{$^2$ EDF R\&D Division Quai Wattiou 78100 Chatou France}

\ead{sergio.chibbaro@upmc.fr }

\begin{abstract}
In this paper, a new formalism for the filtered density function (FDF) approach is developed for
the treatment of turbulent polydispersed two-phase flows in LES simulations. Contrary to
the FDF used for turbulent reactive single-phase flows, the present formalislm is based on 
Lagrangian quantities and, in particular, on the Lagrangian filtered mass density function
(LFMDF) as the central concept. This framework allows modeling and simulation of particle flows for
LES to be set in a rigorous context and various links with other approaches to be made.
In particular, the relation between LES for particle simulations of single-phase flows and
Smoothed Particle Hydrodynamics (SPH) is put forward. Then, the discussion and derivation
of possible subgrid stochastic models used for Lagrangian models in two-phase flows can
set in a clear probabilistic equivalence with the corresponding LFMDF. 
\end{abstract}

\section{Introduction}

Dispersion and entrainment of inertial heavy particles in turbulent flows are crucial in a number of industrial applications and environmental phenomena. Examples are mixing, combustion, depulveration, 
spray dynamics, pollutant dispersion or cloud dynamics. 
Direct numerical simulation (DNS) together with Lagrangian particle tracking (LPT) has been
successfully used to investigate and quantify the behavior of particles in turbulent flow. 
DNS studies have shown, for instance, that particle segregation and subsequent deposition phenomena are related with the interaction between particles and the near-wall coherent vortical structures \citep{Sol_09}. Furthermore, large-scale clustering and preferential concentration, due to 
the tendency of inertial particles to distribute preferentially at the periphery of strong vortical
regions and to segregate into straining regions \citep{Bec_07,Wan_93,Rou_01}, 
have been revealed. 

On the other hand, DNS remains limited to low Reynolds numbers and simple geometries. Conversely,
thanks to the huge growth of available computational resources, large-eddy simulation (LES) can 
address increasingly complex problems. In LES the basic idea is to directly simulate only the turbulence scales larger than a given dimension, while the effects of the unresolved subgrid scales (SGS) on the large-scale motion are modeled\citep{sagaut2006large}. Thus, the exact dynamics of the filtered fluid velocity field can at best be obtained from LES (\textit{ideal LES}), this when SGS modeling and numerical errors are negligible. 
Recent studies \citep{Kue_06,Mar_08} have demonstrated that neglecting the effects of the 
fluid SGS velocity fluctuations on particle motion leads to significant underestimation of preferential concentration and, consequently, to lower deposition fluxes and reduced near-wall accumulation. The issue is therefore how to model the SGS effects in the particle equations.  
Actually, this issue is very similar to the one appearing in the modeling of single-phase turbulent reactive flows\citep{Pop_94} and, in this case, stochastic models that can treat the reactive 
source terms without approximation have shown their great potential. For the same reason, a 
stochastic approach to polydispersed turbulent two-phase flows is interesting.
First ideas of stochastic models have been proposed for LPT coupled with LES of the carrier phase, 
but they have mainly been used for homogeneous isotropic turbulence \citep{Poz_09,Gob_10} and 
they have been proposed with no clear theoretical background.

In this paper, we develop a rigorous probabilistic formalism within which one can develop stochastic 
models for turbulent two-phase flows. As put forward by Pope and coworkers\citep{She_07}, this modelling requires the definition of a proper Filtered density function (FDF). However, at variance with Pope's case, we develop a Lagrangian formalism which includes naturally the Eulerian one. 
In this way, we also show how LES is related to another Lagrangian method, smoothed particle hydrodynamics.

\section{Governing equations}

We consider in this work heavy particles carried by a fluid.
The equations of motion of the fluid are Navier-Stokes equations
\begin{subequations}
\label{fluid: exact field eqs.}
\begin{align}
&\frac{\partial \rho}{\partial t} + \frac{\partial(\rho U_j)}{\partial x_j}=0 \\
&\frac{\partial U_i}{\partial t} + U_j\frac{\partial U_i}{\partial x_j} = 
  -\frac{1}{\rho}\frac{\partial P}{\partial x_i} + 
   \nu \frac{\partial^2 U_i}{\partial x_j^2} \\
&\frac{\partial \Phi_l}{\partial t} + U_j\frac{\partial \Phi_l}{\partial x_j} = 
\Gamma \frac{\partial^2 \Phi_l}{\partial x_j^2} + S_l(\bds{\Phi}) 
\end{align}
\end{subequations}
where the set of scalars $\bds{\Phi}=(m_1, \dots, m_{N_s}, T)$ are the mass fractions 
$m_{l} (l=1, \dots, N_s)$ of the $N_s$ species that compose the reactive mixture and 
the temperature $T$ (if we consider turbulent reactive single-phase flows).
For heavy particles where $\rho_p \gg \rho_f$ transported by the fluid flow, the drag 
and gravity forces are the dominant forces and the particle equation of motion
can be simplified to
\begin{equation}  \label{heavy particle eq}
\frac{d\mb{U}_p}{dt} = \frac{1}{\tau_{p}}(\mb{U}_{s}-\mb{U}_{p})
\;  + \mb{g}.
\end{equation}
The drag force has been written in this form to bring out the particle
relaxation time scale
\begin{equation} \label{definition taup}
\tau_{p}=\frac{\rho_p}{\rho_f}\frac{4\,d_p}{3\,C_D |\mb{U_r}|}.
\end{equation}
The drag coefficient is an
empirical coefficient that can be estimated through experiments.
Various expressions have been put forward\citep{Cli_78}, such as
\begin{equation} \label{expression C_D}
C_D=
\begin{cases}
\displaystyle \frac{24}{Re_p}\left[\,1 + 0.15 Re_p^{0.687}\, \right]
 & \text{if} \; Re_p \leq 1000,  \\
0.44 & \text{if} \; Re_p \geq 1000.
\end{cases}
\end{equation}

\section{Eulerian and Lagrangian descriptions in PDF methods}
\label{PDF methods}

In the case of compressible or reactive flows, 
the proper probability density function (pdf) to be considered is the mass density function $F_L$ 
rather than the pdf itself (which is instead the right choice in the
incompressible case since all fluid particles have a constant mass for a
given volume)\citep{She_07}. Since particles behave like a compresible fluid and may display 
polydispersion, the mass density function is also a relevant choice for two-phase flows\citep{Min_01}.
In this section, we recall some basic definitions (a thorough description can be found elsewhere\citep{Min_01}). 
For a complete description, we introduce the mass density function 
$F_L(t;\mb{y},\mb{V},\bds{\Psi})$ where 
$F_L(t;\mb{y},\mb{V},\bds{\Psi})\,d\mb{y}\,d\mb{V}\,d\bds{\Psi}$ 
represents the probable mass of fluid particles contained in an element
of volume $\,d\mb{y}\,d\mb{V}\,d\bds{\Psi}$ in phase space. The mass
density function is consequently normalized by the total mass of fluid 
\begin{equation} 
M = \int F_L(t;\mb{y},\mb{V},\bds{\Psi}) \,d\mb{y}\,d\mb{V}\,d\bds{\Psi}. 
\end{equation}
These relations are valid both for fluid and solid particles.
The correspondence between the
Eulerian and the Lagrangian descriptions is given by,
\begin{equation}
\label{eq:FL-FE}
F_E(t,\mb{x};\mb{V},\bds{\Psi})= 
F_L(t;\mb{y}=\mb{x},\mb{V},\bds{\Psi})= 
\int F_L(t;\mb{y},\mb{V},\bds{\Psi}) 
\,\delta(\mb{x}-\mb{y})\,d\mb{y},
\end{equation}
where $F_E$ is the Eulerian mass density function.  This relation simply
expresses the fact that, when in phase space $\mb{y}=\mb{x}$, 
there is equivalence
between the two points of view since we consider the same event (in the
incompressible case this relation is verified by $p_E(t,\mb{x};\mb{V})$
and $p_L(t;\mb{y},\mb{V})$). Mass continuity constraint imposes that the integral of $F_E$ over phase space 
$(\mb{V},\bds{\Psi})$ is the expected density at $(t,\mb{x})$ 
(the probable mass of particles in a given state per unit volume). 
The expected density, denoted $\lra{\rho}(t,\mb{x})$, is thus defined by
\begin{equation}
\alpha(t,\mb{x})\,\lra{ \rho }(t,\mb{x})=\int F_E(t,\mb{x};\mb{V},\bds{\Psi})
\,d\mb{V}\,d\bds{\Psi},
\end{equation}

\subsection{PDF theory: No-model equation}

We recall the exact equations for the trajectories of fluid and discrete particles
in a Lagrangian form.
The new set of equations take the following form, 
\begin{equation}
\begin{split}
& dx_{f,i}^+= U_{f,i}^+\, dt, \\
& dU_{f,i}^+= A_{f,i}^+ \, dt + A^+_{p \rightarrow f}(t,\mb{Z},\lra{\mb{Z}})\, dt,\\
& d\Phi_{f,l}^+= \Gamma_f \Delta\Phi_{f,l}^+\, dt +S_f(\bds{\Phi_f^+})\, dt, \\
& dx_{p,i}^+ = U_{p,i}^+ \, dt, \\
& dU_{p,i}^+ = A_{p,i}^+ \, dt, \\
& d\Phi_{p,k}^+= \Gamma_p \Delta\Phi_{p,k}^+\, dt +S_p(\bds{\Phi_p^+})\, dt,
\end{split}
\end{equation} 
where the indexes $l$ and $k$ refer to the dimensions of $\bds{\Phi_f}$
and $\bds{\Phi_p}$, respectively. The $+$
subscript is used to indicate the exact trajectories in contrast to
the modelled ones. Both (exact) accelerations are given by ($A_{f,i}^+$
and $A_{p,i}^+$ are the accelerations of the fluid particles and
the discrete particles, respectively) 
\begin{equation}
\begin{split}
& A_{f,i}^+ = -\frac{1}{\rho_f}
  \frac{\partial P^+}{\partial x_i}+\nu\Delta U_{f,i}^+ , \\
& A_{p,i}^+ =\frac{1}{\tau_{p}} (U_{s,i}^+ -U_{p,i}^+)\; + g_i.
\end{split}
\end{equation} 
A new term is added in the momentum equation of the fluid to account for
the influence of the particles on the fluid. 

Using the techniques developed in stochastic calculus, it is possible to show that the transport equation verified by the Eulerian mass density functions $F_E$ for both phases is of the form (here we show only that for solid particles for the sake of simplcity): 
\begin{eqnarray}
\label{eq:t_p}
\frac{\partial F_E^p}{\partial t}
+  V_{p,i}\frac{\partial F_E^p}{\partial y_{p,i}} &=& 
- \frac{\partial}{\partial V_{p,i}}(A_{p,i}\, F_E^p\,)  \\
&-& \frac{\partial}{\partial V_{s,i}}(\left[A_{s,i}+\lra{A_{p \rightarrow s,i}|\, \mb{y}_p,\mb{V}_p,\bds{\Psi}_p}\right]\, F_E^p\,)\\
              &-&\frac{\partial}{\partial\Psi_{p,k}}(\lra{\Gamma_p\Delta\Phi_{p,k}^+\,|\,\mb{Z}=\mb{z}}\, F_E^p\,) -\frac{\partial}{\partial \Psi_{p,k}}(\,S_{p,k}(\bds{\Psi_p}^+)\,F_E^p\,)\notag,
\end{eqnarray}

\section{The LES approach for single-phase turbulent flows}

In single-phase flows, the starting point is provided by the Navier-Stokes equations 
supplemented with conservation equations for a set of scalars.
These are field equations (the different fields are density $\rho(t,\bf{x})$, 
pressure $P(t,\mb{x})$, velocity $\mb{U}(t,\mb{x})$ and scalars 
$\bds{\Phi}(t,\mb{x})$, where $\mb{x}$ represents the coordinates in physical space). 
The LES approach consists in calculating the larger three-dimensional unsteady turbulent 
motions while the effects of the smaller scale motions are modelled. For that purpose,
LES involves the use of a spatial filter
\begin{equation}
\lra{f({\mb x},t)}_L=\int_{-\infty}^{\infty} f({\bf y},t)G({\bf y},{\bf x})d{\bf y}
\end{equation}
where $G$ is the filter function, $f (x,t)_L$ represents the
filtered value of the transport variable $f (x,t)$, and $f^{\prime}=f-\lra{f} _L$
denotes the fluctuations of f from the filtered value. We consider spatially and temporally 
invariant and localized filter functions, thus $G (y,x)\equiv G(x-y)$ with the
properties, $G(x)=G(-x)$, and $\int G(x)dx=1$.

Starting from Navier-Stokes equations, application of the filtering operator yields:
 \begin{eqnarray}
 \label{eq:les}
 \frac{\partial \lra{U_j}_L}{\partial x_j}&=&0 \\
\frac{\partial \lra{U_i}_L}{\partial t}&+&\lra{U_j}_L\frac{\partial \lra{U_i}_L}{\partial x_j}=
-\frac{1}{\rho}\frac{\partial \lra{P}_L}{\partial x_i} +\nu\Delta \lra{ U_i}_L 
-\frac{\partial \lra{\tau_{ij}}_L}{\partial x_j},\\
\frac{\partial \lra{\Phi_l}_L}{\partial t}&+&\lra{U_j}_L\frac{\partial \lra{\Phi_l}_L}{\partial x_j}=
\Gamma\Delta \lra{ \Phi_l}_L 
-\frac{\partial \lra{M^l_{j}}_L}{\partial x_j}+\lra{S_l}_,
\end{eqnarray}
where $\lra{\tau_{ij}}_L=\lra{U_i U_j}_L- \lra{U_i}_L \lra{U_j}_L$
and $\lra{M^l_{j}}_L=\lra{\Phi_l U_j}_L - \lra{\Phi_l}_L \lra{U_j}_L$
denote the subgrid stress and the subgrid mass flux respectively.

\subsection{The Filtered Density Function (FDF) approach} \label{single-phase FDF}

In this section, we recall the main points of the FDF theory introduced for reactive flows, 
following the approach proposed by Pope and developed with coworkers\citep{She_07}. For
general compressible single-phase flows, the FDF theory introduces a "filtered mass density 
function" (FMDF), denoted $F$, 
\begin{equation}
\label{eq:pl}
F(t,\mb{x};\bds{V},\bds{\Psi})= 
\int \rho(\bds{U}({\bf  y},t),\bds{\Phi}({\bf  y},t),\bds{V},\bds{\Psi}) G({\bf y}-{\bf x})d{\bf y}~,
\end{equation}
\begin{equation}
\rho(\bds{U}({\bf  y},t),\bds{\Phi}({\bf  y},t),\bds{V},\bds{\Psi})=
\delta(\bds{V}-\bds{U}({\bf  y},t)) \otimes \delta(\bds{\Psi}-\bds{\Phi}({\bf  y},t))~,
\end{equation}
where $\delta$ denotes the delta function and $\Psi$ denotes the composition
domain of the scalar array. The term $\rho(\bds{U}({\bf  y},t),\bds{\Phi}({\bf  y},t)$
is the fine-grained mass density and Eq. ~(\ref{eq:pl}) implies that the FMDF is the 
spatially filtered value of the fine-grained mass density. Thus, $F$ gives the density in the 
composition space of the fluid around x weighted by the filter G. With the
condition of a positive filter kernel, $F$ has all the properties of a PDF. 
It is also useful to define the conditional filtered value of a variable $Q({\bf  x},t)$:
\begin{equation}
\lra{Q(\mb{x},t)\, \vert \, \bds{V},\bds{\Psi})}_L\equiv
\frac{\int Q(\mb{y},t)\rho(\bds{U}({\bf  y},t),\bds{\Phi}({\bf  y},t))G({\bf y}-{\bf x})d{\bf y}}{F(t,\mb{x};\bds{V},\bds{\Psi})}~,
\end{equation}
which implies the important integral property:
$$\lra{Q(\mb{x},t)}_L =
\int \lra{Q(\mb{x},t)\, \vert \, \bds{\Psi}}_L F(t,\mb{x};\bds{V},\bds{\Psi}) d\bds{V}d\bds{\Psi}.$$

By standard techniques, it is possible to show that the governing equation for the fdf is
\begin{equation}
\label{eq:fdf}
\frac{\partial F}{\partial t} + \frac{\partial [V F]}{\partial y_i} =
-\frac{\partial}{\partial V_i}
[\, \lra{A_{f,i}\,|\,\bds{V},\bds{\Psi}}\,F \,]
-\frac{\partial}{\partial \Psi_l}
[\, \lra{ \Gamma \Delta \Phi_l \,|\, \bds{V},\bds{\Psi}}\,F \,]
-\frac{\partial}{\partial \Psi_l}[\,S_l(\bds{\Psi})\,F \,].
\end{equation}
The important aspect is that only an Eulerian FMDF is considered and that, apart from
the numerical Monte-Carlo particles used later on the theory~\citep{She_07}, there is
no notion of a Lagrangian filtered density.

\section{A new FDF formalism for polydispersed two-phase flows}

In turbulent reactive single-flows, the FDF formalism developed so far by Pope and co-workers
is based only on the Eulerian filtered density function, which is consistent with the fact that
the basic exact equations to handle are field equations. However, in polydispersed two-phase
flows, the physical situation is completely different since the basic
exact equations (for the particle phase) are Lagrangian. Therefore, in order to develop
a proper LES treatment for particle flows, it is not possible to copy simply what has been
achieved in single-phase flows. A new point of view must be introduced and a new formalism
must be entirely developed. 

The theory briefly recalled in Section~\ref{PDF methods} has shown that the key notion is 
the Lagrangian Mass Density Function. For the development of a FDF approach to two-phase flows, 
we choose therefore to retain a Lagrangian point of view as the starting point and
we introduce as the central concept the \textit{Lagrangian filtered mass density function} 
(LFMDF) which is defined as (considering, for sake of simplicity, only a particle phase)
\begin{equation}
\label{eq:Ftild}
\widetilde{F}_L^p(t;\mb{y},\bds{V},\bds{\Psi})= 
\int F_L^p(t;{\bf  y'},\bds{V},\bds{\Psi}) G({\bf y}-{\bf y'})d{\bf y'}~,
\end{equation}
In terms of fine-grained densities, this can be written for the description of 
the particle phase (noted $F_L^p$) with $N$ the number of particles
present in the domain at the time $t$ as: 
\begin{eqnarray}
\label{eq:Ftild2}
\widetilde{F}^{p}_L(t;\mb{y},\bds{V},\bds{\Psi})&= &
\int \sum_{i=1}^N  m_{p,i}\, G({\bf y}-{\bf y'})\delta(\bds{y'}-\bds{x}_i(t))\otimes
\delta(\bds{V}-\bds{U}_i(t)) \otimes \delta(\bds{\Psi}-\bds{\Phi}_i(t))d{\bf y'}\notag \\
&=&\sum_{i=1}^N m_{p,i}\, G({\bf y}-{\bf x_i}(t))\otimes \delta_V \otimes \delta_{\Psi}~.
\end{eqnarray}
In order to avoid a confusion between the filtering operator and the Eulerian and Lagrangian
indices (E and L, respectively) which will have to be introduced, the filtering operator
of a variable $A$ will be noted by $\widetilde{A}$ (instead of $\lra{A}_L$ in previous sections).

For further developments, it is useful to consider the conditional filtered value of a variable 
$H(t)$ which is defined as
\begin{eqnarray}
\label{eq:condfilt}
\widetilde{ \left(H |{\bf  y},\bds{V},\bds{\Psi}\right) } \widetilde{F}^{p}_L&=&
\int  H(t) {F_L^p}(t;{\bf  y'},\bds{V},\bds{\Psi}) G({\bf y}-{\bf y'})d{\bf y'}~,\notag \\
&=&\sum_{i=1}^N H_i   m_{p,i} G({\bf y}-{\bf x_i})\otimes \delta_V \otimes \delta_{\Psi}~.
\end{eqnarray}
As in Section~\ref{single-phase FDF}, this operator has all the properties of a conditional
average and a key property is that for any function of the variables in the state-vector,
$H=H(\bds{x}, \bds{U}, \bds{\Phi})$ 
\be
\widetilde{\left(H |{\bf y},\bds{V},\bds{\Psi}\right) }= H({\bf y},\bds{V},\bds{\Psi}).
\ee
In this formalism, following what is done in PDF methods, we choose to define the 
Eulerian filtered mass density function (EFMDF) directly from the LFMDF as
\be
\label{eq:FE1}
 \widetilde{F}_E^p(t,\mb{x};\mb{V},\mb{\Psi}) =\, \widetilde{F}_L^p(t;\mb{y}=\mb{x},\mb{V}_f,\mb{\Psi}).
\ee
From these definitions, we have thus that
\be
 \widetilde{F}_E^p(t,\mb{x};\mb{V},\mb{\Psi})  =\, 
 \int {F_L^p}(t;{\bf  y'},\bds{V},\bds{\Psi}) G({\bf x}-{\bf y'})d{\bf y'}~.
\ee
or in terms of the fine-grained densities
\be
 \widetilde{F}_E^p(t,\mb{x};\mb{V},\mb{\Psi})  =\,
 \sum_{i=1}^N m_{p,i}\, G({\bf x}-{\bf x_i}(t))\otimes \delta_V \otimes \delta_{\Psi}.
\ee
From the EFMDF, we can define the filtered for any variable $Q(t,\mb{x})$
\begin{equation} \label{Eulerian Filtered field}
\alpha_{p}(t,\mb{x})\lra{\rho}_p \widetilde{Q} (t,\mb{x}) =
\int \int \widetilde{\left(Q |{\bf  y},\bds{V},\bds{\Psi}\right) }
\;\; \widetilde{F}_E^p(t,\mb{x};\mb{V},\bds{\psi})
\,d\mb{V}\,d\bds{\psi}
\end{equation}
and for a function of the variables in the state-vector
\begin{equation} 
\alpha_{p}(t,\mb{x})\lra{\rho}_p \widetilde{Q} (t,\mb{x}) =
\int \int Q(t, \mb{x}, \mb{V}, \bds{\psi})
\;\; \widetilde{F}_E^p(t,\mb{x};\mb{V},\bds{\psi})
\,d\mb{V}\,d\bds{\psi}
\end{equation}
where $\alpha_{p}(t,\mb{x})\lra{\rho}_p$ is the filtered local value of the particle mass 
fraction at $t$ and position $\mb{x}$ 
\begin{equation}
\alpha_{k}(t,\mb{x}) \lra{\rho}_p=
\int \widetilde{F}_E^p(t,\mb{x};\mb{V}_k,\bds{\psi}_k)
\,d\mb{V}_p\,d\bds{\psi}_k.
\end{equation}

Given the filtered mass density function, it is possible to derive the associated evolution equation 
using the fine-grained definition (\ref{eq:Ftild2}) with standard techniques \cite{Pop_85}
(considering in the following derivation that $m_{p,i}$ are constant, for the sake of simplicity)
\begin{align}
\frac{\partial \widetilde{F}^{p}_L}{\partial t} &= 
\sum_{i=1}^N \left(m_{p,i} \frac{\partial G}{\partial t}\delta_{V \Psi}+m_{p,i} G \frac{\partial \delta_V}{\partial t}\delta_{\Psi}+
m_{p,i} G \frac{\partial \delta_{\Psi}}{\partial t}\delta_{V}\right) \notag \\
&=\sum_{i=1}^N \left(m_{p,i} \frac{ \partial G}{\partial \mathbf{x}}\frac{d \mathbf{x}_i}{d t}
\delta_{V \Psi}- m_{p,i} G \frac{d \mathbf{U}_i}{d t}\frac{\partial \delta_V}{\partial \mathbf{V}}\delta_{\Psi}-
m_{p,i} G \frac{d \mathbf{\Phi}_i}{d t} \frac{\partial \delta_{\Psi}}{\partial \bf{\Psi}} \delta_{{V}} \right) \notag \\
&=\sum_{i=1}^N 
\left(-m_{p,i} \frac{ \partial G}{\partial \mathbf{y}}\frac{d \mathbf{x}_i}{d t}\delta_{V \Psi}-
m_{p,i} G \frac{d \mathbf{U}_i}{d t}\frac{\partial \delta_V}{\partial \mathbf{V}}\delta_{\Psi}-
m_{p,i} G \frac{d \mathbf{\Phi}_i}{d t} \frac{\partial \delta_{\Psi}}{\partial \bf{\Psi}} \delta_{{V}} \right) \notag \\
&=\sum_{i=1}^N 
\left( -\frac{ \partial }{\partial \mathbf{y}}(m_{p,i}G \frac{d \mathbf{x}_i}{d t}\delta_{V \Psi})-
\frac{\partial}{\partial \mathbf{V}}(m_{p,i} G \frac{d \mathbf{U}_i}{d t} \delta_V \delta_{\Psi})-
\frac{\partial }{\partial \bf{\Psi}}(m_{p,i} G \frac{d \mathbf{\Phi}_i}{d t}  \delta_{{V}}\delta_{\Psi}) \right) \notag \\
&=-\frac{ \partial }{\partial \mathbf{y}}
[\, \widetilde{\left( \frac{d \mathbf{x}}{d t} |{\bf  y},\bds{y},\bds{\Psi}\right)} \,
\widetilde{F}_L^p \, ] 
-\frac{\partial}{\partial \mathbf{V}}
[\, \widetilde{\left( \frac{d \mathbf{U}}{d t} |{\bf  y},\bds{V},\bds{\Psi}\right) } \widetilde{F}_L^p \,]
-\frac{\partial }{\partial  \Psi}
[\, \widetilde{\left( \frac{d\mathbf{\Phi}}{d t} | {\bf  y},\bds{V},\bds{\Psi}\right)}
\widetilde{F}_L^p \, ] \notag \\
&=-\frac{\partial [\, \mathbf{V} \widetilde{F}_L^p \,]}{\partial \mathbf{y}}
-\frac{\partial }{\partial \mathbf{V}} 
[\, \widetilde{\left(\mathbf{A}_u |{\bf y},\bds{V},\bds{\Psi}\right)} \widetilde{F}_L^p \,]
-\frac{\partial }{\partial \Psi}[\, \widetilde{\left( \mathbf{A}_{\Phi}| {\bf y},\bds{V},\bds{\Psi}\right)}
\widetilde{F}_L^p \,]
\end{align}
This equation corresponds to the system of Lagrangian equations which are written, retaining
general notations since we will apply them for fluid and solid particles in following
sections, as: 
\begin{align}
dx_{i} &= U_{i}\, dt, \\
dU_{i} &= A_{u,i} \, dt \\
d\Phi_{l}&=A_{\phi,l} \, dt, 
\end{align}
Since the EFMDF is by definition identical to the LFMDF, the same equation is satisfied
automatically by $\widetilde{F}_E^p$, from which any transport equations for filtered fields
can be easily extracted by integration over the corresponding sample-space, 
cf. Eq.~(\ref{Eulerian Filtered field}). Then, by decomposising into filtered and "fluctuations" 
(or unresolved values), the equation for the FMDFs can be expressed as
\begin{align}
\frac{\partial \widetilde{F}^{p}_L}{\partial t} +
\frac{\partial [\, \widetilde{\mb{U}} \widetilde{F}_L^p \,]}{\partial \mathbf{y}} =&
-\frac{\partial }{\partial \mb{V}} [\, \widetilde{\mb{A}_u}\, \widetilde{F}_L^p \,]
-\frac{\partial }{\partial \Psi}[\, \widetilde{\mathbf{A}_{\Phi}} \widetilde{F}_L^p \,] \notag \\
& - \frac{\partial}{\partial \mathbf{y}}
[\, \left(\mathbf{V} - \widetilde{\mb{U}}\right) \widetilde{F}_L^p \,] \notag \\
& -\frac{\partial }{\partial \mathbf{V}} 
[\, \left\{ \widetilde{\left(\mathbf{A}_u |{\bf y},\bds{V},\bds{\Psi}\right)} 
- \widetilde{\mb{A}_u} \right\} \widetilde{F}_L^p \,] \\
&-\frac{\partial }{\partial \Psi}
[\, \left\{ \widetilde{\left( \mathbf{A}_{\Phi}| {\bf y},\bds{V},\bds{\Psi}\right)}
- \widetilde{\mathbf{A}_{\Phi}} \right\} \widetilde{F}_L^p \,] \notag
\end{align}
where the first line corresponds to the effects of resolved scales (closed terms) and the 
last three ones to the effects of the unresolved modes (unclosed terms).

\section{Smoothed Particle Hydrodynamics (SPH) as a particle-based LES}

As a first application of the present general formalism, we consider the case of 
$N$ fluid particles (in which case, we will use here an index $f$ instead of $p$).
If we disregard the effects of unresolved scales, the transport equation for
the EFMDF is simply:
\be
\frac{\partial \widetilde{F}^{f}_L}{\partial t} +
\frac{\partial [\, \widetilde{\mb{U}} \widetilde{F}_L^f \,]}{\partial \mathbf{y}} =
-\frac{\partial }{\partial \mb{V}} [\, \widetilde{\mb{A}_u}\, \tilde{F}_L^f \,]
\ee
The particle (or Monte-Carlo) solution of this equation is 
\begin{align}
& d\widetilde{x}_{f,i}= \widetilde{U}_{f,i}\, dt, \\
& d\widetilde{U}_{f,i}= \widetilde{A}_{f,i} \, dt 
\end{align} 
where $A_{f,i}$ is the Navier-Stokes equation in a Lagrangian formulation
\be
\label{eq:Af}
{A}_{f,i}= -\frac{1}{\rho_f}\frac{\partial P}{\partial x_i}+\nu\Delta U_{f,i}.
\ee
By applying the general relations for a particle labelled $a$ (and using label $b$ 
to express the sum over other particles), we have that
\be
\alpha_a \rho_a \widetilde{A}_a = \sum_b m_b A_b W_{ab}
\ee
where we have used for the filtering function, $G$ a kernel function, $W$, with a 
typical length $h$
\be
G({\bf y}_a-{\bf y}_b)=W({\bf y}_a-{\bf y}_b, h)=W_{ab}.
\ee
For incompressible flows, the density is constant $\rho_a=\rho_b$ and the local
mass fraction estimation gives (writing $m_b= \rho_b\, \mathcal{V}_b$ with 
$\mathcal{V}_b$ the particle volume)
\be
\alpha_a \rho_a = \sum_b m_b W_{ab}= \sum_b \rho_b \mathcal{V}_b W_{ab}
\;\,, \text{and that} \;\, \alpha_a=\sum_b \mathcal{V}_b W_{ab}=1~,
\ee
which is the normalisation condition for the kernel $W$ and states that the domain
is continuously filled with fluid particles. Thus, we obtain the following particle
equations
\begin{align}
& d\widetilde{x}_{f,i}= \widetilde{U}_{f,i}\, dt, \\
& d\widetilde{U}_{f,i}= \sum_b 
\left[ \left( -\frac{1}{\rho_{f}} \frac{\partial P}{\partial x_i}\right)_b+\left(\nu\Delta U_{f,i}\right)_b\right] \mathcal{V}_b W_{ab} \, dt 
\end{align} 

Therefore, by disregarding unresolved effects and considering a Lagrangian description
of a single-phase fluid flow, we obtain exactly the smooth particle hydrodynamics (SPH) formalism\citep{Mon_05}. Further SPH expressions, in particular for truly 
incompressible flows~\citep{Sze_11}, give the complete equations by expressing the 
pressure-gradient and viscous terms with a particle formulation:
\begin{align}
& \frac{d\widetilde{\mb{x}}_{a}}{dt}= \widetilde{\mb{U}}_{a}, \\
& \frac{d\widetilde{\mb{U}}_{a}}{dt}= 
\sum_b m_b \left\{ \frac{\widetilde{p}_a + \widetilde{p}_b}{\rho_a\, \rho_b} 
+ 8 \frac{\nu_a + \nu_b}{\rho_a + \rho_b}\frac{ \widetilde{\mb{U}_{ab}}
\cdot\mb{r}_{ab}}{r_{ab}^2} \right\}
\nabla_a W_{ab}
\end{align} 

Thus, as a first conclusion, it is now rigorously demonstrated that the SPH method is 
equivalent to performing a Lagrangian LES without a subgrid model. This is a rather intuitive
notion, considering the smoothing nature of the SPH operator, but there was 
previously no theoretical framework to express it in a strict setting. In other words, 
present SPH formulations appears a particle-based LES without any sub-kernel model.
And, consequently, a direct extension could be to consider sub-kernel models, for
example with the form of particle stochastic models, which would provide a direct 
link with particle simulations in LES. 

\section{LES for particle flows and subgrid stochastic models}

We now consider polydispersed solid particles embedded in a turbulent described
by a LES simulation. Following the discussion and choice made in PDF methods~\cite{Min_01},
it appears interesting to include the velocity of the fluid seen by the particles
in the state-vector retained and, thus, to handle the LFMDF
$\widetilde{F}_L^p(t; \mb{y}_p,\mb{V}_p,\mb{V}_s)$. The transport equation for
the LFMDF (and EFMDF $\widetilde{F}_E^p(t, \mb{y}_p; \mb{V}_p,\mb{V}_s)$) is then 
\begin{equation}
\frac{\partial \widetilde{F}^{p}_L}{\partial t} +
\frac{\partial [\, \mathbf{V}_p \widetilde{F}_L^p \,]}{\partial \mathbf{y}_p} =
-\frac{\partial }{\partial \mathbf{V}_p} 
[\, \left( \frac{\mb{U}_s - \mb{U}_p}{\tau_p} \right) \widetilde{F}_L^p \,]
-\frac{\partial }{\partial \mathbf{V}_s} 
[\, \widetilde{\left(\mathbf{A}_s |{\bf y}_p,\bds{V}_p,\bds{V}_s\right)} \widetilde{F}_L^p \,]
\end{equation}
which corresponds to the particle equations
\begin{align}
d\mb{x}_p &= \mb{U}_p\, dt \notag \\
d\mb{U}_p &= \frac{\mb{U}_s - \mb{U}_p}{\tau_p}\, dt \\
d\mb{U}_s &= \mb{A}_s\, dt.
\end{align}
where a model for $\mb{A}_s$ has to be proposed. 

The first idea that was applied in LES simulations with solid particles was to disregard
subgrid effects and to identify the (instantaneous) fluid velocity seen, $\mb{U}_s(t)$,
with the available filtered value at the particle location, 
$\mb{U}_s(t)=\widetilde{\mb{U}_s}(t, \mb{x}_p(t))$. This amounts to considering only particle
location and velocity in the corresponding LFMDF since $\mb{U}_s(t)$ is now given,
$\widetilde{F}_p(t; \mb{y}_p,\overline{\mb{V}}_p)$ whose evolution equation is
\begin{equation}
\frac{\partial \widetilde{F}^{p}_L}{\partial t} +
\frac{\partial [\, \overline{\mb{V}}_p \widetilde{F}_L^p \,]}{\partial \mathbf{y}_p} =
-\frac{\partial }{\partial \overline{\mb{V}}_p} 
[\, \left( \frac{\widetilde{\mb{U}_s} - \mb{U}_p}{\widetilde{\tau_p}} \right) \widetilde{F}_L^p \,].
\end{equation}
This equation involves only first-order derivatives and is an example of a classical Liouville
equation (note that the notation for the sample value of particle velocity, $\mb{U}_p$, has
been changed to $\overline{\mb{V}}_p$ to indicate that, in this case, particle velocitues are
only driven by a known drag force). Indeed, the Montec-Carlo solution of this transport equation
corresponds simply to the deterministic particle equations of motion
\begin{align}
d\widetilde{\mb{x}}_{p} &= \widetilde{\mb{U}_{p}}\, dt, \\
d\widetilde{\mb{U}}_{p} &= \frac{\widetilde{\mb{U}_s} - \widetilde{\mb{U}_p}}{\widetilde{\tau_p}}\, dt
\end{align}
where $\widetilde{\tau_p}$ represents the particle relaxation timescale expressed only in 
terms of the filtered relative velocity $\widetilde{\mb{U}_r}$. 

However, in order to take into account the explicit effects of unresolved scales, the next step
is to introduce a stochastic model for $\mb{U}_s$ which, in the corresponding sample-space
for the LFMDF $\widetilde{F}_L^p(t; \mb{y}_p,\mb{V}_p,\mb{V}_s)$ means that we have now 
a modeled transport equation, for example a Fokker-Planck equation
\begin{equation}
\frac{\partial \widetilde{F}^{p}_L}{\partial t} +
\frac{\partial [\, V_{p,i} \widetilde{F}_L^p \,]}{\partial y_{p,i}} = 
-\frac{\partial}{\partial V_{p,i}}
[\, \left( \frac{U_{s,i} - U_{p,i}}{\tau_p} \right) \widetilde{F}_L^p \,]  
-\frac{\partial [\, D_{s,i}  \widetilde{F}_L^p \,]}{\partial V_{s,i}} 
+\frac{1}{2}\frac{\partial^2 [\, B^2_{s,ij} \widetilde{F}_L^p \,]}{\partial^2 V_{s,i}V_{s,j}} 
\end{equation}
which corresponds to a stochastic diffusion process for the modeled particle variables
\begin{align}
d\mb{x}_p &= \mb{U}_p\, dt \notag \\
d\mb{U}_p &= \frac{\mb{U}_s - \mb{U}_p}{\tau_p}\, dt \\
d\mb{U}_s &= \mb{D}_s\, dt + \underline{\mb{B}}_s\, d\mb{W}
\end{align}
with $d\mb{W}$ the increments of the vector of independent Wiener processes. For practical
examples of precise diffusion stochastic processes (stochastic modeling being not 
necessarily limited to this subclass!), it is possible to consider the extensions of
the stochastic models used in PDF models~\citep{Min_04}. For example, a simple model would
be 
\begin{align}
& dx_{p,i}= U_{p,i}\, dt \\
& dU_{p,i} = \frac{U_{s,i} - U_{p,i}}{\tau_p}\, dt \\
& dU_{s,i} = -\frac{1}{\rho_f}\frac{\partial \widetilde{P}}{\partial x_i}\, dt
+ \left( \widetilde{U_{p,j}} - \widetilde{U_{f,j}} \right) 
\frac{\partial \widetilde{U_f}}{\partial x_j}\, dt
- \frac{U_{s,i} - \widetilde{U_{s,i}}}{\tau_{L,i}^{*}}\, dt 
+ \sqrt{C_i^{*} \widetilde{\epsilon}}\, dW_i
\end{align}
where crossing-trajectory effects have been kept for the timescales of the unresolved
fluid velocity seen $\tau_{L,i}^{*}$ (with $\tau_L$ the timescale of the fluid unresolved
motions). It is thus important to note that
the diffusion matrix, $\underline{\mb{B}}_s$, is diagonal but not isotropic. In LES simulations,
the fluid subgrid fluctuations can be regarded as nearly isotropic (as a first step) and
the general expression proposed in PDF models for $C_i^{*}$ would be simplified to
\begin{equation}
C_i^{*}= C_0\, b_i + \frac{2}{3}(b_i -1) \quad \text{with} \quad b_i=\frac{\tau_L}{\tau_{L,i}^{*}}
\end{equation}
The development of the corresponding filtered particle fields, as well as a more detailed
discussion of several possible subgrid kernels, will be presented in forthcoming manuscripts.
Yet, the present framework already emphasizes that it is crucial to consider satisfactory 
diffusion coefficients (or diffusion matrix) for the velocity of the fluid seen. If some
arguments have been put forward on the form of the drift vector, $\mb{D}_s$, an essential
aspect remains an acceptable form of $\underline{\mb{B}}_s$ so as to reproduce, at least, a 
correct energy flux from the resolved to unresolved scales. Note that, the simple
idea to use a closure expression as in single-phase flows, that is 
$B_{s,ij}= \sqrt{C_0\, \widetilde{\epsilon}}\, \delta_{ij}$, is \textit{drastically false}
since a physically-incorrect and spurious energy flux, or transfer rate, is induced
inconsistent with the modeled rate $\widetilde{\epsilon}$. 

\section{Conclusions}
In this paper, a theoretical formalism for the LES of particle turbulent polydispersed two-phase flows has been presented. The formalism has been developed for the Lagrangian FDF, for which the general transport equation is derived. In this framework, it is possible to build  stochastic models for particle flows in a rigorous way. Furthermore, we have applied this formalism to show that SPH is a particle-based LES. Finally, a first example of Langevin model constructed within the formalism is proposed considering isotropic sub-grid fluctuations.


\bibliographystyle{jfm}
\bibliography{etc}

\end{document}